\documentclass[amsmath,amssymb,aps,pra,amsfonts,superscriptaddress,showkeys,showpacs,conference,12pt]{revtex4}
\usepackage{amsmath}
\usepackage{graphicx,amsmath}
\usepackage{booktabs}
\usepackage{amssymb}
\usepackage{graphicx}
\usepackage{subfigure}
\usepackage{phonetic}
\usepackage{extarrows}
\usepackage{float}
\usepackage{amsfonts}
\usepackage{epsfig}
\usepackage{bm}
\usepackage{mathrsfs}
\usepackage{bbm}
\usepackage{dsfont}
\usepackage{multirow}
\usepackage{makecell}
\usepackage{textcomp}
\usepackage{float}
\usepackage{color}
\usepackage[colorlinks,citecolor=blue,linkcolor=blue]{hyperref}
\hypersetup{urlcolor=blue}

\date{\today}

\begin{document}
\title{Qubit-assisted squeezing of the mirror motion in a dissipative optomechanical cavity system}
\author{Cheng-Hua Bai}
\affiliation{Department of Physics, Harbin Institute of Technology, Harbin, Heilongjiang 150001, China}
\author{Dong-Yang Wang}
\affiliation{Department of Physics, Harbin Institute of Technology, Harbin, Heilongjiang 150001, China}
\author{Shou Zhang}
\email{szhang@ybu.edu.cn}
\affiliation{Department of Physics, Harbin Institute of Technology, Harbin, Heilongjiang 150001, China}
\affiliation{Department of Physics, College of Science, Yanbian University, Yanji, Jilin 133002, China}
\author{Hong-Fu Wang}
\email{hfwang@ybu.edu.cn}
\affiliation{Department of Physics, College of Science, Yanbian University, Yanji, Jilin 133002, China}

\begin{abstract}
We investigate a hybrid system consisting of an atomic ensemble trapped inside a dissipative optomechanical cavity assisted with the perturbative oscillator-qubit coupling. It is shown that such a hybrid system is very suitable for generating stationary squeezing of the mirror motion in the long-time limit under the unresolved sideband regime. Based on the approaches of master equation and covariance matrix, we discuss the respective squeezing effects in detail and find that in both approaches, simplifying the system dynamics
with adiabatic elimination of the highly dissipative cavity mode is very effective. In the approach of master equation, we find that the squeezing is a resulting effect of cooling process and is robust against the thermal fluctuations of the mechanical mode. While in the covariance matrix approach, we can obtain the analytical result of the steady-state mechanical position variance from the reduced dynamical equation approximately. Finally, we compare the two different approaches and find that they are completely equivalent for the stationary dynamics. The scheme may be meaningful for the possible ultraprecise quantum measurement involved mechanical squeezing.

\end{abstract}
\maketitle

\section{Introduction}\label{Sec1}
Many significant progresses have been achieved with the recent advance of cavity optomechanics over the last few years~\cite{2014RMP861391}. Examples include ground-state cooling of the mechanical mode~\cite{2007PRL99093901,2007PRL99093902,2014PRA90053841,2015SC58516,2018PRA98023816}, macroscopic entanglement between two spatially separated movable mirrors~\cite{2014PRA89023843,2015SC58518,2018FP13130319}, optical multistability behavior~\cite{2016PRA93023844,2017SC60010311}, and so on.

Thereinto, generation of non-classical states of motion around the ground state based on the cavity optomechanical system is one of the most effective methods to study the quantum effects at mesoscopic or macroscopic scales. Specifically, the quantum squeezing associated with the mechanical motion, reduction of the quantum fluctuation in its position or momentum below the quantum noise limit, is not only of significant importance for testing the quantum fundamental theory~\cite{2012PT6529}, such as exploring the quantum-classical boundary~\cite{1991PT4436}, but also has widely potential applications, such as the detection of gravitational waves~\cite{1980RMP52341,1992Science256325,1999PT5244}. Thus, achieving squeezed state in mechanical oscillator (mirror) is a greatly desired goal.

To this end, several well-known methods and techniques to generate squeezing of the mechanical mode were proposed~\cite{1991PRL673665,2009PRL103213603,2011PRA83033820,2013OE21020423,2018OE26013783,2013PRA88063833,2009PRA79063819,2010PRA82033811,2016PRA93043844,2015PRA91013834,2010PRA82021806R,2014PRA89023849,2018PRA97043619,2018PRA98023807,2008APL92133102,2015PRL115243601}. One of the early most outstanding schemes was to modulate the frequency of the oscillator~\cite{1991PRL673665}. Nevertheless, although this is simplest, it is not easy to utilize for many different types of mechanical systems. Subsequently, the alternative methods based on the cavity optomechanical system to overcome this drawback have been proposed. Examples include modulation of the external driving laser~\cite{2009PRL103213603,2011PRA83033820,2013OE21020423,2018OE26013783}; adoption of one red detuned and the other blue detuned two-zone driving sources~\cite{2013PRA88063833}; direct squeezing transfer from the squeezed external driving field or squeezed cavity field generated by the parametric amplifier inside the cavity to the oscillator~\cite{2009PRA79063819,2010PRA82033811,2016PRA93043844}; use of the Duffing nonlinearity~\cite{2015PRA91013834}, etc. While concentrating on the linear radiation pressure interaction, the squeezing of the mechanical mode in quadratically coupled optomechanical system has also been investigated. In this case, one can drive the cavity with two beams~\cite{2010PRA82021806R} and use the bang-bang control technique to kick the mechanical mode~\cite{2014PRA89023849}. Meanwhile, we have noted very recently that the effects of the cooperations between the squeezed field driving and quadratic optomechanical coupling~\cite{2018PRA97043619} and between the periodically modulated driving and parametric driving~\cite{2018PRA98023807} on the generation of mechanical squeezing are investigated. The stronger mechanical squeezing can be viewed as the joint effect in the cooperation regime.

In fact, the basic mechanism for creating mechanical squeezing is to introduce a parametric coupling for the motional degree of freedom of the oscillator. The Hamiltonian takes the form $H\propto b^2+b^{\dag2}$ (where $b$ and $b^{\dag}$ are the annihilation and creation operators of the oscillator) and the corresponding evolution operator is a squeezed operator so that the squeezing can be achieved effectively~\cite{QuantumOptics}. Therefore, a significantly interesting question is how the parametric coupling can be reached in cavity optomechanical system. Very fortunately, we have noted that this type of parametric coupling has been used to enhance the quantum correlations in optomechanical system and it can be introduced by perturbatively coupling a single qubit to the mechanical oscillator~\cite{2018AOP39239}. In addition, the photon blockade and two-color optomechanically induced transparency in this kind of model have been discussed in detail~\cite{2015PRA92033806,2014PRA90023817}. Meanwhile, the oscillator-qubit coupling can also be realized in experiments successfully based on the superconducting quantum circuit system~\cite{2009Nature459960,2018PRA98023821}.

On the other hand, as we all know, the master equation is a powerful tool to study the evolution of a practical quantum system dynamics in quantum theory~\cite{QuantumOptics}. However, since the dynamics of fluctuations is linearized and the noises are Gaussian in general optomechanical system, it is greatly convenient to introduce the covariance matrix to study the system dynamics~\cite{2009PRL103213603,2007PRL98030405,2014PRA89023843}. But to our knowledge, the dynamical results obtained from the two different approaches have not been compared until now.

In this paper, we study the squeezing effect of mechanical oscillator induced by the oscillator-qubit coupling in a hybrid system consisting of an atomic ensemble trapped inside a dissipative optomechanical cavity. We discuss the mechanical squeezing in detail based on the approaches of master equation and covariance matrix, respectively. In the approach of master equation, we eliminate the highly dissipative cavity mode adiabatically and obtain the effective Hamiltonian. By solving the master equation numerically, we find that the steady-state squeezing of mechanical oscillator can be generated successfully in the long-time limit. We also demonstrate that the squeezing is the resulting effect of cooling process. By numerically and dynamically deriving the optimal effective detuning simultaneously, we check the cooling effects when the mechanical oscillator is prepared in a thermal state with certain mean thermal phonon number initially.

As to the approach of covariance matrix, by eliminating the highly dissipative cavity mode adiabatically, the dynamical equation of $6\times6$ covariance matrix can be reduced as the one of $4\times4$ covariance matrix, which significantly simplifies the system dynamics. In the appropriate parameter regime, the analytical solution of the steady-state variance for the oscillator position can be obtained approximately. Finally, we make a clear comparison for these two different approaches. We find that the steady-state dynamics in the long-time limit obtained from the two different approaches are completely equivalent.

This paper is organized as follows. In Sec.~\ref{Sec2}, we introduce the hybrid system model under consideration and derive the Hamiltonian of the system. In Sec.~\ref{Sec3}, we discuss the squeezing effect of mechanical oscillator in detail based on the approaches of master equation and covariance matrix, respectively. In Sec.~\ref{Sec4}, we give a brief discussion about the implementation of present scheme with the circuit-QED system. Finally, a conclusion is given in Sec.~\ref{Sec5}.

\section{System and model}\label{Sec2}
\begin{figure}
\centering
\includegraphics[scale=0.8]{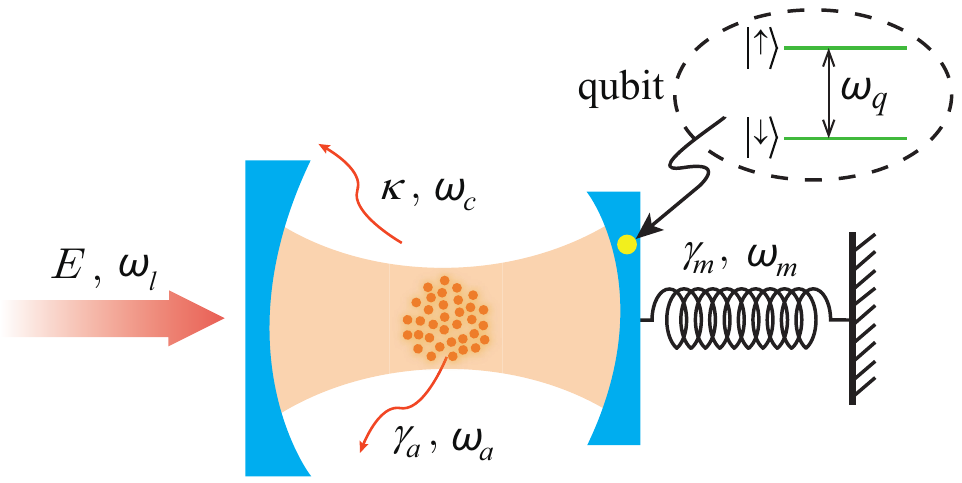}
\caption{(Color online) Schematic diagram of the considered system. A cloud of identical two-level atoms is trapped in a dissipative optomechanical cavity, which is driven by an external laser field. The qubit (within the black dashed elliptical ring) which is denoted by a yellow dot inside the movable mirror has the levels $|\uparrow\rangle$ and $|\downarrow\rangle$.}\label{Fig1}
\end{figure}

The system under consideration is schematically shown in Fig.~\ref{Fig1},  where a cloud of identical two-level atoms (with frequency $\omega_a$ and decay rate $\gamma_a$) is trapped in a dissipative optomechanical cavity (with frequency $\omega_c$ and decay rate $\kappa$). An external laser field with time-independent amplitude $E$ and frequency $\omega_l$ drives the optomechanical cavity and the movable mirror coupled with a qubit is modeled as the mechanical oscillator with frequency $\omega_m$ and damping rate $\gamma_m$. The mechanical oscillator is coupled to the cavity field via the radiation-pressure interaction. The Hamilton of the system is given by (in the unit of $\hbar=1$)
\begin{eqnarray}\label{Eq1}
H&=&\omega_ca^{\dag}a+\omega_aS_z+\frac{\omega_m}{2}(q^2+p^2)+2\eta q^2+ \cr\cr
&&g(S_+a+S_-a^{\dag})-g_0a^{\dag}aq+E(a^{\dag}e^{-i\omega_lt}+ae^{i\omega_lt}),
\end{eqnarray}
in which $a$ ($a^{\dag}$) is the annihilation (creation) operator of the cavity field, $S_{+,-,z}=\sum\limits_i\sigma^{(i)}_{+,-,z}$ are the collective spin Pauli operators of atoms, and $q$ ($p$) is the dimensionless position (momentum) operator of the mechanical oscillator, satisfying the standard canonical commutation relation $[q, p]=1$. $g$ and $g_0$ represent, respectively, the atom-cavity coupling strength and the single-photon radiation-pressure coupling strength. In Hamiltonian Eq.~(\ref{Eq1}), the first three terms in first line correspond to the free Hamiltonian of the driven cavity, atoms, and mechanical oscillator, respectively. The fourth term refers to the Hamiltonian for the qubit-oscillator interaction, where $2\eta$ is the coupling strength. As to the generation of this term, we will make a discussion finally. The first two terms in second line describe the coupling between atoms and cavity field and the optomechanical interaction between the cavity field and mechanical oscillator, respectively. The last term gives the driving of the cavity by an external laser field.

The spin operators of the atoms can be described in terms of a collective bosonic operator, $c=S_-/\sqrt{N}$. For the sufficiently large atom number $N$ and weak atom-cavity couping, $S_z\simeq c^{\dag}c-N/2$~\cite{2015PRA92033841}. In the rotating frame with respect to laser frequency $\omega_l$, the Hamiltonian can be rewritten as
\begin{eqnarray}\label{Eq2}
H^{\prime}=\delta_ca^{\dag}a+\Delta_ac^{\dag}c+\frac{\omega_m}{2}(q^2+p^2)+2\eta q^2+G(c^{\dag}a+ca^{\dag})-g_0a^{\dag}aq+E(a^{\dag}+a),
\end{eqnarray}
where $\delta_c=\omega_c-\omega_l$ and $\Delta_a=\omega_a-\omega_l$ are, respectively, the cavity and atomic detuning with respect to the external driving laser. $G=\sqrt{N}g$ is the collective atom-cavity coupling strength.

In the following, we will discuss the squeezing effect of the movable mirror in detail based on the approaches of master equation and covariance matrix, respectively.

\section{Discussion of the squeezing for the movable mirror}\label{Sec3}
\subsection{The approach of master equation}
\subsubsection{Hamiltonian}
To discuss the squeezing of the movable mirror based on the approach of master equation, it is better to introduce the annihilation and creation operators of the mechanical oscillator
\begin{eqnarray}\label{Eq3}
b=(q+ip)/\sqrt{2},~~~~~~~~~~b^{\dag}=(q-ip)/\sqrt{2}.
\end{eqnarray}
In terms of $b$ and $b^{\dag}$, the Hamiltonian in Eq.~(\ref{Eq2}) can be rewritten as
\begin{eqnarray}\label{Eq4}
H^{\prime\prime}=\delta_ca^{\dag}a+\Delta_ac^{\dag}c+\omega_m^{\prime}b^{\dag}b+\eta(b^2+b^{\dag2})+G(c^{\dag}a+ca^{\dag})-g_0^{\prime}a^{\dag}a(b+b^{\dag})+E(a^{\dag}+a),
\end{eqnarray}
where $\omega_m^{\prime}=\omega_m+2\eta$ and $g_0^{\prime}=g_0/\sqrt{2}$.

In general, besides the coherent dynamics, the quantum systems will also be inevitably coupled to their environments. Taking all the damping and noise effects into account, the evolution of the system can be completely described by the following nonlinear quantum Langevin equations (QLEs)
\begin{eqnarray}\label{Eq5}
\frac{da}{dt}&=&-(\kappa+i\delta_c)a-iGc+ig_0^{\prime}a(b+b^{\dag})-iE+\sqrt{2\kappa}a_{\mathrm{in}}(t), \cr\cr
\frac{db}{dt}&=&-(\gamma_m+i\omega_m^{\prime})b-2i\eta b^{\dag}+ig_0^{\prime}a^{\dag}a+\sqrt{2\gamma_m}b_{\mathrm{in}}(t), \cr\cr
\frac{dc}{dt}&=&-(\gamma_a+i\Delta_a)c-iGa+\sqrt{2\gamma_a}c_{\mathrm{in}}(t),
\end{eqnarray}
where $a_{\mathrm{in}}(t)$, $b_{\mathrm{in}}(t)$, and $c_{\mathrm{in}}(t)$ are the noise operators for the cavity field, mechanical oscillator, and atoms, respectively, which have zero mean value and satisfy the following correlation functions
\begin{eqnarray}\label{Eq6}
\langle a_{\mathrm{in}}(t)a_{\mathrm{in}}^{\dag}(t^{\prime})\rangle&=&\delta(t-t^{\prime}), ~~~~~~~~~~~~~~~~~~~~
\langle a_{\mathrm{in}}^{\dag}(t)a_{\mathrm{in}}(t^{\prime})=0, \cr\cr
\langle b_{\mathrm{in}}(t)b_{\mathrm{in}}^{\dag}(t^{\prime})\rangle&=&(n_m+1)\delta(t-t^{\prime}), ~~~~~~~~
\langle b_{\mathrm{in}}^{\dag}(t)b_{\mathrm{in}}(t^{\prime})=n_m\delta(t-t^{\prime}), \cr\cr
\langle c_{\mathrm{in}}(t)c_{\mathrm{in}}^{\dag}(t^{\prime})\rangle&=&\delta(t-t^{\prime}), ~~~~~~~~~~~~~~~~~~~~
\langle c_{\mathrm{in}}^{\dag}(t)c_{\mathrm{in}}(t^{\prime})=0,
\end{eqnarray}
in which $n_m=\left[\mathrm{exp}(\hbar\omega_m/k_BT)-1\right]^{-1}$ is the mean thermal phonon number. Here $T$ is the environment temperature of mechanical reservoir and $k_B$ is the Boltzmann constant.

The strong driving on the cavity leads to the large amplitudes for the cavity field, mechanical mode, and atoms. Thus, the standard linearization procedure can be applied to simplify the dynamical equations. To this end, we express the operators in Eq.~(\ref{Eq5}) as the sum of their mean values and quantum fluctuations, i.e., $\mathscr{O}(t)\rightarrow\langle\mathscr{O}(t)\rangle+\mathscr{O}(t)~(\mathscr{O}=a, b, c)$. Hence, the dynamical equation corresponding to the mean values is given by the following set of nonlinear differential equations:
\begin{eqnarray}\label{Eq7}
\langle\dot{a}(t)\rangle&=&-(\kappa+i\delta_c)\langle a(t)\rangle-iG\langle c(t)\rangle+ig_0^{\prime}\langle a(t)\rangle(\langle b(t)\rangle+\langle b(t)\rangle^{\ast})-iE, \cr\cr
\langle\dot{b}(t)\rangle&=&-(\gamma_m+i\omega_m^{\prime})\langle b(t)\rangle-2i\eta\langle b(t)\rangle^{\ast}+ig_0^{\prime}|\langle a(t)\rangle|^2, \cr\cr
\langle\dot{c}(t)\rangle&=&-(\gamma_a+i\Delta_a)\langle c(t)\rangle-iG\langle a(t)\rangle.
\end{eqnarray}	
On the other hand, the dynamics of the quantum fluctuations is governed by the following linearized QLEs:
\begin{eqnarray}\label{Eq8}
\dot{a}&=&-(\kappa+i\delta_c)a-iGc+ig_0^{\prime}\langle a(t)\rangle(b+b^{\dag})+ig_0^{\prime}(\langle b(t)\rangle+\langle b(t)\rangle^{\ast})a+\sqrt{2\kappa}a_{\mathrm{in}}(t), \cr\cr
\dot{b}&=&-(\gamma_m+i\omega_m^{\prime})b-2i\eta b^{\dag}+ig_0^{\prime}\langle a(t)\rangle^{\ast}a+ig_0^{\prime}\langle a(t)\rangle a^{\dag}+\sqrt{2\gamma_m}b_{\mathrm{in}}(t), \cr\cr
\dot{c}&=&-(\gamma_a+i\Delta_a)c-iGa+\sqrt{2\gamma_a}c_{\mathrm{in}}(t).
\end{eqnarray}

\begin{figure}
\centering
\includegraphics[scale=0.5]{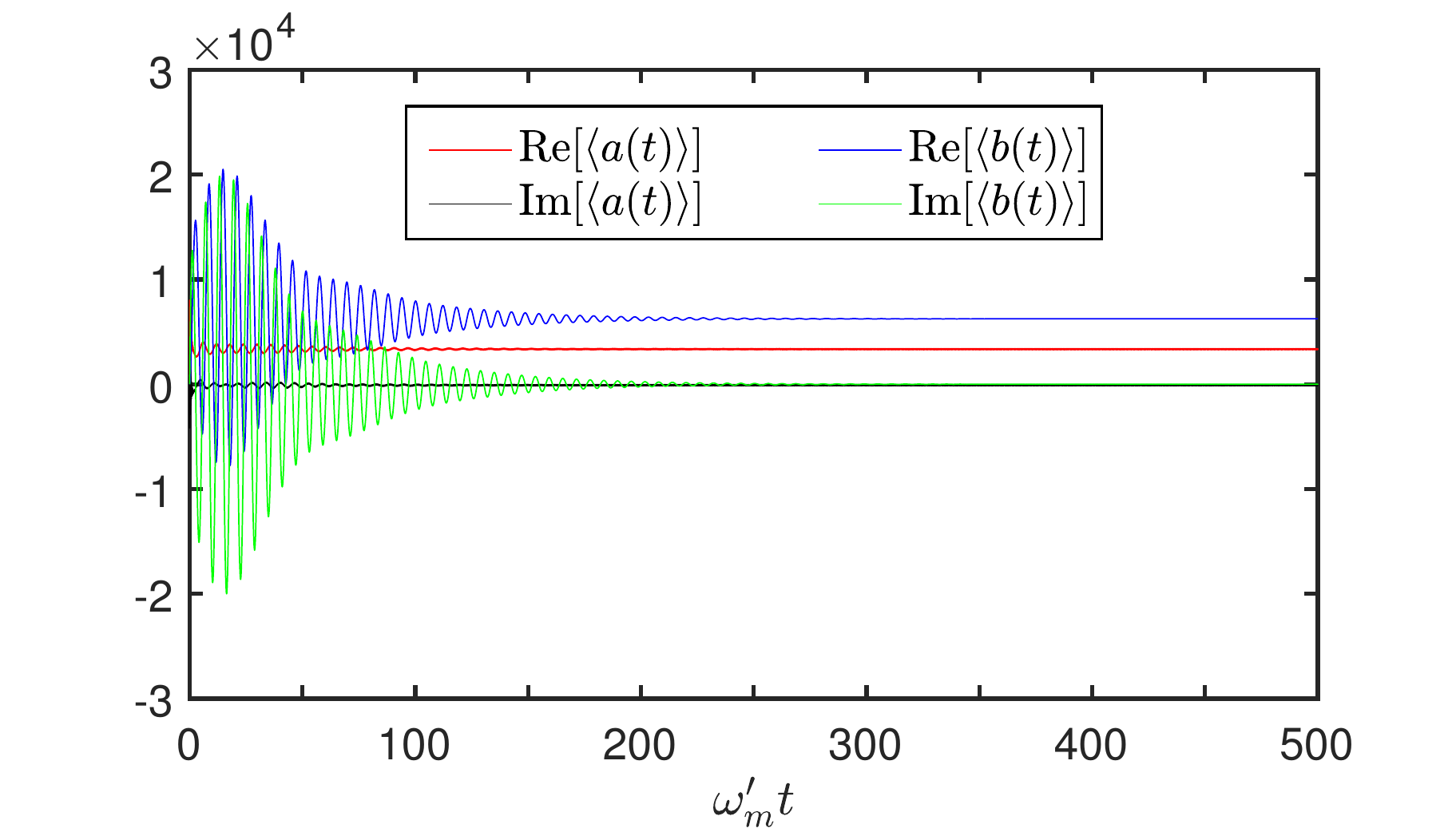}
\caption{(Color online) Time evolution of the real and imaginary parts of the cavity mode mean value $\langle a(t)\rangle$ and the mechanical mode mean value $\langle b(t)\rangle$.
The system parameters are chosen as: $\omega_m=\pi\times10^6~\mathrm{Hz}$, $\gamma_m=10^{-6}\omega_m$, $g_0^{\prime}=10^{-3}\omega_m$, $\omega_c=10^8\omega_m$, $\delta_c=-250\omega_m$, $\kappa=3\omega_m$, $\Delta_a=1.1\omega_m$, $\gamma_a=0.1\omega_m$, $G=8\omega_m$, $\eta=0.2\omega_m$, $P=20~\mathrm{mW}$, and $E=\sqrt{2P\kappa/(\hbar\omega_l)}$.}\label{Fig2}
\end{figure}

Via solving Eq.~(\ref{Eq7}) numerically, we plot the time evolution of the real and imaginary parts of the cavity mode mean value $\langle a(t)\rangle$ and the mechanical mode mean value $\langle b(t)\rangle$ in Fig.~\ref{Fig2}. From Fig.~\ref{Fig2}, we can find that the real and imaginary parts of the mean values reach their steady states quickly and the real part is much larger than the imaginary part ($\mathrm{Re}[\langle a(t)\rangle]\gg\mathrm{Im}[\langle a(t)\rangle]$ and $\mathrm{Re}[\langle b(t)\rangle]\gg\mathrm{Im}[\langle b(t)\rangle]$). As a consequence, we can make the following approximations safely:
\begin{eqnarray}\label{Eq9}
\langle a(t)\rangle\simeq\langle a(t)\rangle^{\ast}\simeq|\langle a\rangle_s|,~~~~~~~~~~
\langle b(t)\rangle\simeq\langle b(t)\rangle^{\ast}\simeq|\langle b\rangle_s|,
\end{eqnarray}
where $\langle a\rangle_s$ and $\langle b\rangle_s$ represent, respectively, the steady state mean values of cavity mode and mechanical mode.

So the Hamiltonian of the system for the quantum fluctuations corresponding to Eq.~(\ref{Eq8}) can be written as
\begin{eqnarray}\label{Eq10}
H_{\mathrm{lin}}=\Delta_ca^{\dag}a+\omega_m^{\prime}b^{\dag}b+\Delta_ac^{\dag}c+\eta(b^2+b^{\dag2})+G(c^{\dag}a+ca^{\dag})-G_0(a+a^{\dag})(b+b^{\dag}),
\end{eqnarray}
in which $\Delta_c=\delta_c-2g_0^{\prime}|\langle b\rangle_s|$ is the effective cavity detuning and $G_0=g_0^{\prime}|\langle a\rangle_s|$ is the effective optomechanical coupling strength.

Under the parameter regimes $|\Delta_c|\gg(\omega_m^{\prime}, |\Delta_a|)$ and $\kappa\gg(\gamma_m, \gamma_a)$, the cavity mode can be eliminated adiabatically and the solution of the fluctuation operator $a(t)$ at the time scale $t\gg1/\kappa$ can be obtain (see Appendix \ref{App1})
\begin{eqnarray}\label{Eq11}
a(t)\simeq\frac{iG_0[b(t)+b^{\dag}(t)]}{\kappa+i\Delta_c}+\frac{-iGc(t)}{\kappa+i\Delta_c}+A_{\mathrm{in}}(t),
\end{eqnarray}
where $A_{\mathrm{in}}(t)$ is the modified noise operator.

Substituting Eq.~(\ref{Eq11}) into the expressions about modes $b$ and $c$ in Eq.~(\ref{Eq8}), we obtain the QLEs about $b$ and $c$ after eliminating cavity mode $a$ adiabatically
\begin{eqnarray}\label{Eq12}
\dot{b}&\simeq&-(\gamma_m+i\tilde{\omega}_m)b-iG_{\mathrm{eff}}(c+c^{\dag})-2i\eta^{\prime}b^{\dag}+b^{\prime}_{\mathrm{in}}(t), \cr\cr
\dot{c}&\simeq&-(\gamma_{\mathrm{eff}}+i\Delta_{\mathrm{eff}})c-iG_{\mathrm{eff}}(b+b^{\dag})+c_{\mathrm{in}}^{\prime}(t),
\end{eqnarray}
where $b_{\mathrm{in}}^{\prime}(t)$ and $c_{\mathrm{in}}^{\prime}(t)$ represent the modified noise terms. The effective parameters for mechanical frequency $\tilde{\omega}_m$, optomechanical coupling $G_{\mathrm{eff}}$, bilinear strength $\eta^{\prime}$, detuning $\Delta_{\mathrm{eff}}$, and decay rate $\gamma_{\mathrm{eff}}$ are defined as, respectively,
\begin{eqnarray}\label{Eq13}
\tilde{\omega}_m&=&\omega_m^{\prime}-\frac{2G_0^2\Delta_c}{\Delta_c^2+\kappa^2}=\omega_m+2\eta^{\prime},~~~~~~~~~~
G_{\mathrm{eff}}=\left|\frac{G_0G}{\Delta_c-i\kappa}\right|, \cr\cr
\eta^{\prime}&=&\eta-\frac{G_0^2\Delta_c}{\Delta_c^2+\kappa^2},~~~~~~~
\Delta_{\mathrm{eff}}=\Delta_a-\frac{G^2\Delta_c}{\Delta_c^2+\kappa^2},~~~~~~~
\gamma_{\mathrm{eff}}=\gamma_a+\frac{G^2\kappa}{\Delta_c^2+\kappa^2}.
\end{eqnarray}

Therefore, the effective Hamiltonian corresponding to QLEs about mechanical mode $b$ and atom mode $c$ in Eq.~(\ref{Eq12}) is
\begin{eqnarray}\label{Eq14}
H_{\mathrm{eff}}=\tilde{\omega}_mb^{\dag}b+\Delta_{\mathrm{eff}}c^{\dag}c+G_{\mathrm{eff}}(b+b^{\dag})(c+c^{\dag})+\eta^{\prime}(b^{\dag2}+b^2).
\end{eqnarray}

\subsubsection{Generation of the mechanical squeezing}
We now introduce the quadrature operators for the mechanical mode $X=(b+b^{\dag})/\sqrt{2}$ and $Y=(b-b^{\dag})/\sqrt{2}i$, so the variance of the quadrature operator $Z~(Z=X, Y)$ is determined by
\begin{eqnarray}\label{Eq15}
\langle\delta Z^2\rangle=\langle Z^2\rangle-\langle Z\rangle^2=\mathrm{Tr}[Z^2\varrho(t)]-\mathrm{Tr}[Z\varrho(t)]^2,
\end{eqnarray}
where $\varrho(t)$ is the system density operator, the dynamics of which is completely governed by the following master equation
\begin{eqnarray}\label{Eq16}
\dot{\varrho}(t)=-i[H_{\mathrm{lin}}, \varrho]+\kappa\mathcal{D}[a]\varrho+\gamma_m(n_m+1)\mathcal{D}[b]\varrho+\gamma_mn_m\mathcal{D}[b^{\dag}]\varrho+
\gamma_a\mathcal{D}[c]\varrho,
\end{eqnarray}
in which $\mathcal{D}[o]\varrho=o\varrho o^{\dag}-(o^{\dag}o\varrho+\varrho o^{\dag}o)/2~(o=a, b, c)$ is the standard Lindblad superoperators.

According to the Heisenberg uncertainty principle, the product of the variances $\langle\delta X^2\rangle$ and $\langle\delta Y^2\rangle$ satisfies the following inequality,
\begin{eqnarray}\label{Eq17}
\langle\delta X^2\rangle\langle\delta Y^2\rangle\geq|\frac12[X, Y]|^2,
\end{eqnarray}
where $[X, Y]=i$. Thus if either $\langle\delta X^2\rangle$ or $\langle\delta Y^2\rangle$ is below $1/2$, the state of the movable mirror exhibits the behavior of quadrature squeezing.

\begin{figure}
\centering
\includegraphics[scale=0.5]{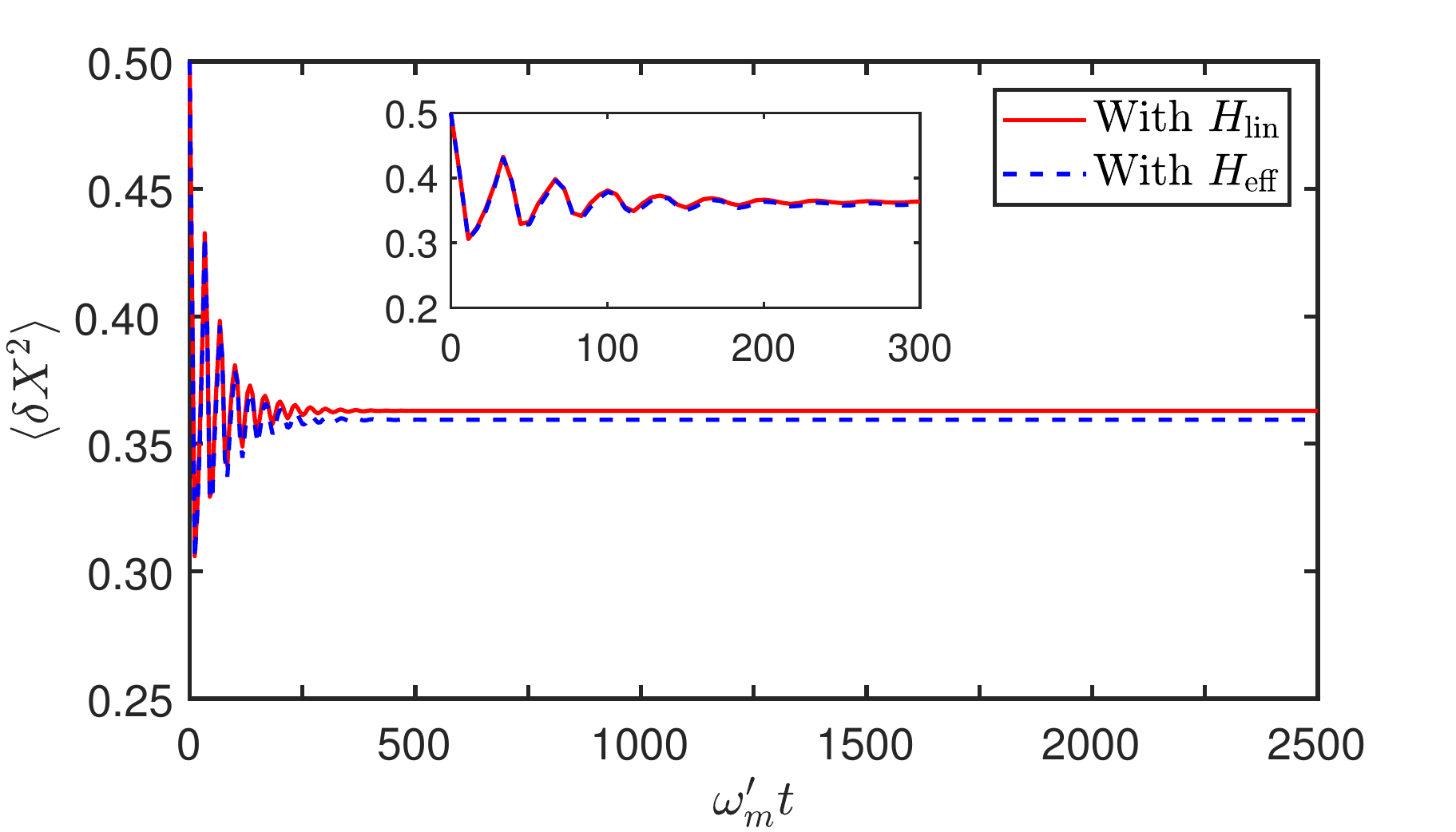}
\caption{(Color online) Time evolution of the variance $\langle\delta X^2\rangle$ for the quadrature operator $X$. The red solid line and blue dashed line denote, respectively, the corresponding numerical result with original linearized Hamiltonian $H_{\mathrm{lin}}$ and effective Hamiltonian $H_{\mathrm{eff}}$. The system parameters are presented in Fig.~\ref{Fig2}. Here the mean thermal phonon number has been set $n_m=0$.}\label{Fig3}
\end{figure}

In Fig.~\ref{Fig3}, we present the time evolution of the variance $\langle\delta X^2\rangle$ for the quadrature operator $X$ with the original linearized Hamiltonian in Eq.~(\ref{Eq10}). We find that the variance $\langle\delta X^2\rangle$ finally converges to a steady-state value below $1/2$ after the transitory oscillation. Moreover, to check the validity for the adiabatic elimination of cavity mode $a$, it is very necessary to solve the effective master equation
\begin{eqnarray}\label{Eq18}
\dot{\varrho}_{\mathrm{eff}}(t)=-i[H_{\mathrm{eff}}, \varrho_{\mathrm{eff}}]+
\gamma_m(n_m+1)\mathcal{D}[b]\varrho_{\mathrm{eff}}+\gamma_mn_m\mathcal{D}[b^{\dag}]\varrho_{\mathrm{eff}}+\gamma_{\mathrm{eff}}\mathcal{D}[c]\varrho_{\mathrm{eff}},
\end{eqnarray}
where $\gamma_{\mathrm{eff}}$ is the effective decay rate of atoms, as given in Eq.~(\ref{Eq13}). We also give the time evolution of the variance $\langle\delta X^2\rangle$ in this case in Fig.~\ref{Fig3} and find that the numerical results obtained from the original linearized Hamiltonian $H_{\mathrm{lin}}$ and effective Hamiltonian $H_{\mathrm{eff}}$ agree well. Thus, in the present parameter regime, simplifying the system dynamics with adiabatic elimination of cavity mode is valid.

\begin{figure}
\centering
\includegraphics[scale=0.5]{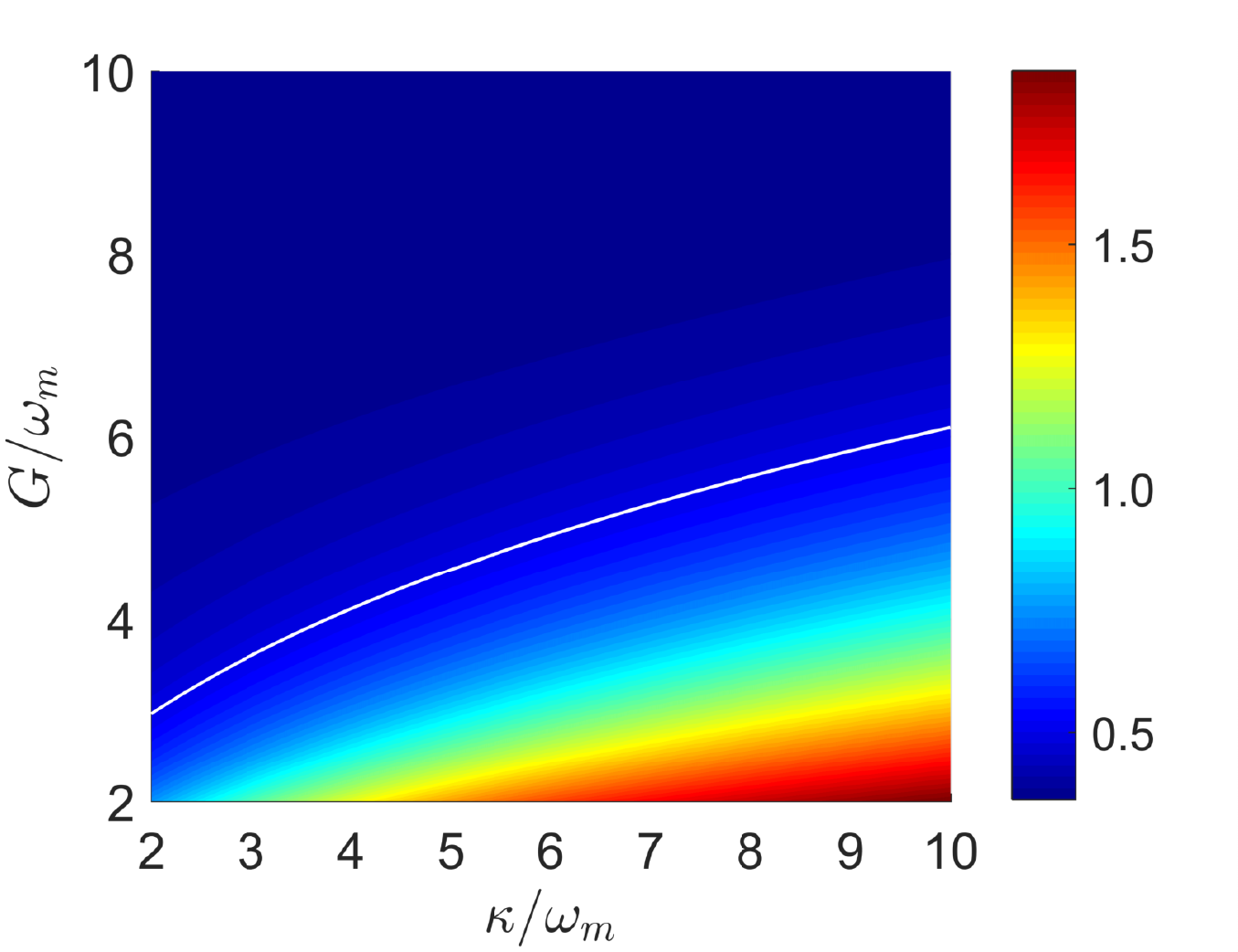}
\caption{(Color online) Steady-state variance $\langle\delta X^2\rangle$ for the quadrature operator $X$ versus the cavity decay rate $\kappa$ and atom-cavity coupling strength $G$ with mean thermal phonon number $n_m=0$. Here the parameters are the same as those in Fig.~\ref{Fig2} and the white curve denotes the contour line of quantum noise limit.}\label{Fig4}
\end{figure}

Next, we present steady-state variance $\langle\delta X^2\rangle$ for the quadrature operator $X$ versus the cavity decay rate $\kappa$ and atom-cavity coupling strength $G$ in Fig.~\ref{Fig4}. One notes that the squeezing of movable mirror can be generated successfully even in the high dissipative optomechanical cavity $(\kappa>\omega_m)$, as long as the coupling strength $G$ is appropriate. This originates from the strong enough atom-cavity coupling effectively suppresses the undesired effect of cavity dissipation on the mechanical squeezing.

\subsubsection{The optimal effective detuning $\Delta_{\mathrm{eff}}$}
In Eq.~(\ref{Eq14}), the last term describes a parametric-amplification process and plays the paramount role in the generation of squeezing. While the third term describes an effective optomechanical coupling process that leads to cooling and heating of the mechanical mode simultaneously. As is well known, to reveal the quantum effects including mechanical squeezing at the macroscopic level, it is a prerequisite to suppress the heating process as soon as possible.

\begin{figure}[H]
\centering
\includegraphics[scale=0.5]{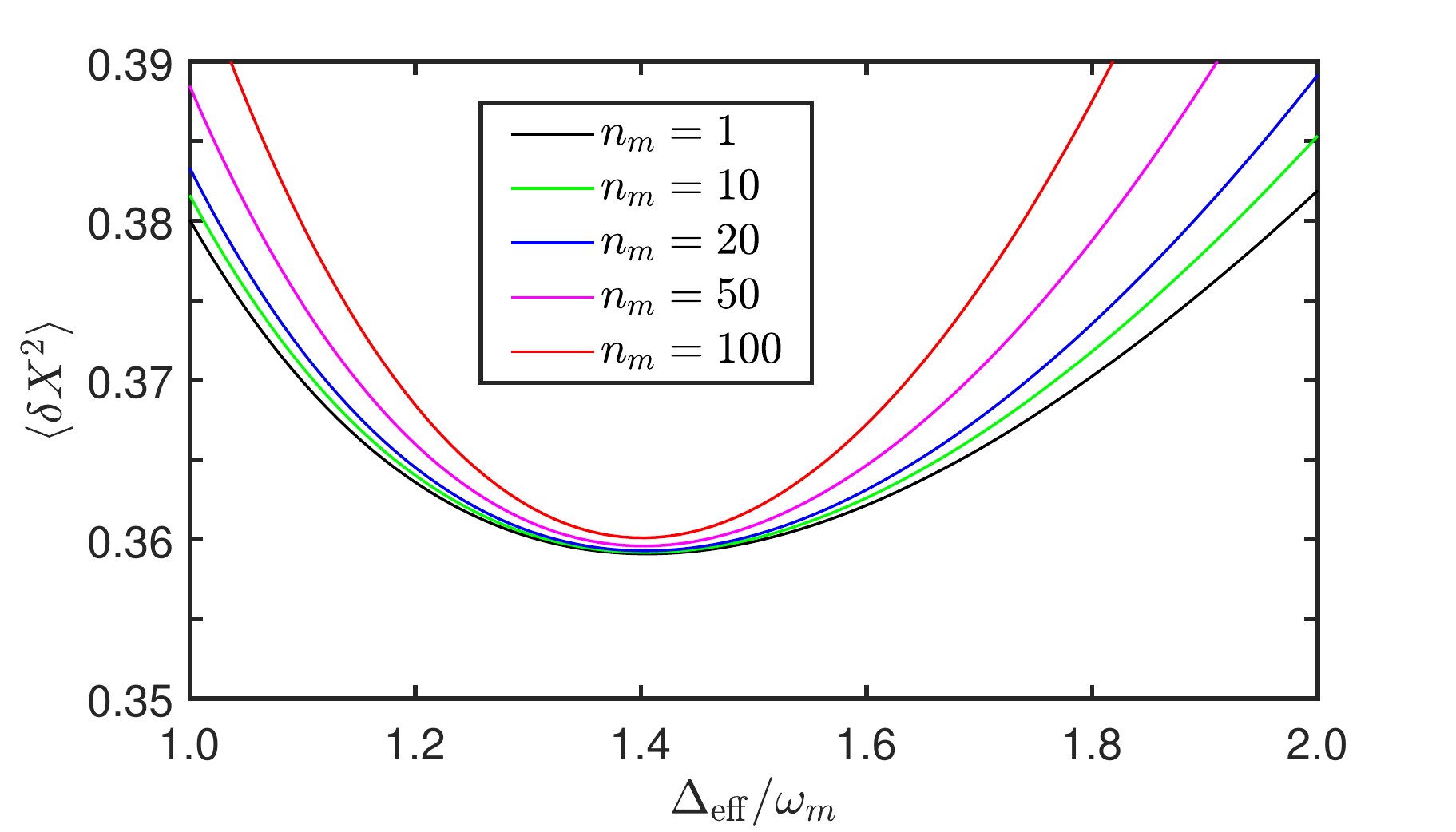}
\caption{(Color online) The dependence of steady-state variance $\langle\delta X^2\rangle$ for the quadrature operator $X$ on the effective detuning $\Delta_{\mathrm{eff}}$ in the case of different mean thermal phonon numbers. The other parameters are fixed as in Fig.~\ref{Fig2}.}\label{Fig5}
\end{figure}

In Fig.~\ref{Fig5}, we present the dependence of steady-state variance $\langle\delta X^2\rangle$ for the quadrature operator $X$ on the effective detuning $\Delta_{\mathrm{eff}}$ in the case of different mean thermal numbers. We find that, as is expected, the more mean thermal phonon number exists, the larger steady-state variance $\langle\delta X^2\rangle$ will become, but there is an optimal effective detuning  point $\Delta_{\mathrm{eff}}\simeq1.4\omega_m$. This is because at this point, the heating process of mechanical mode is strongly suppressed. Thus, the destructive effect of thermal noises on the squeezing of movable mirror is almost non-existent. Next, to give more insight of the physical mechanism, we analyze the optimal effective detuning $\Delta_{\mathrm{eff}}$ from the system dynamics.

\begin{figure}
\centering
\includegraphics[scale=0.8]{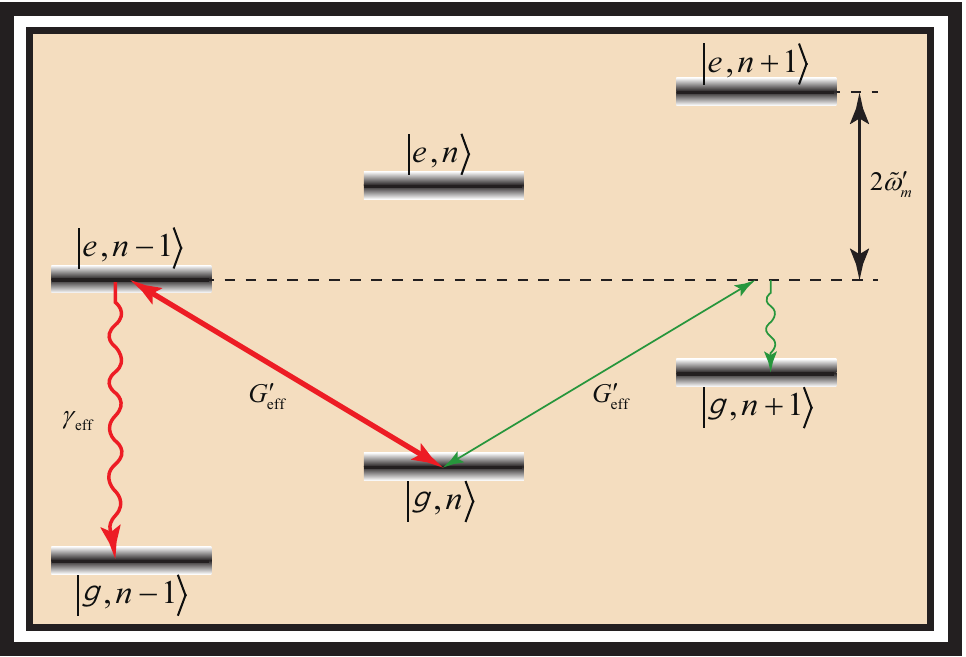}
\caption{(Color online) Energy-level diagram of the transformed Hamiltonian in Eq.~(\ref{Eq23}). Here the product state $|g, n\rangle$ ($|e, n\rangle$) denotes the atomic ground state $|g\rangle$ (excited state $|e\rangle$) and $n$ mechanical phonons. The red arrows represent the cooling mechanism corresponding to the anti-Stokes process in the resonant condition $\Delta_{\mathrm{eff}}=\tilde{\omega}_m^{\prime}$ while the green arrows represent the heating mechanism corresponding to the Stokes process.}\label{Fig6}
\end{figure}

We first apply the squeezing transformation $S(r)=\mathrm{exp}\left[\frac r2(b^2-b^{\dag2})\right]$ with the squeezing parameter (see Appendix \ref{App2})
\begin{eqnarray}\label{Eq19}
r=\frac14\ln\left(1+\frac{4\eta^{\prime}}{\omega_m}\right),
\end{eqnarray}
to the effective Hamiltonian $H_{\mathrm{eff}}$ in Eq.~(\ref{Eq14}). In this transformation, 
\begin{eqnarray}\label{Eq20}
S^{\dag}(r)bS(r)=\cosh rb-\sinh rb^{\dag}, ~~~~~~~~~~
S^{\dag}(r)cS(r)=c.
\end{eqnarray}
The transformed effective Hamiltonian is thus given by
\begin{eqnarray}\label{Eq21}
H_{\mathrm{eff}}^{\prime}=S^{\dag}(r)H_{\mathrm{eff}}S(r)
=\tilde{\omega}_m^{\prime}b^{\dag}b+\Delta_{\mathrm{eff}}c^{\dag}c+G_{\mathrm{eff}}^{\prime}(b+b^{\dag})(c+c^{\dag}),
\end{eqnarray}
where
\begin{eqnarray}\label{Eq22}
\tilde{\omega}_m^{\prime}=\omega_m\sqrt{1+\frac{4\eta^{\prime}}{\omega_m}},~~~~~~~~~~
G_{\mathrm{eff}}^{\prime}=G_{\mathrm{eff}}\left(1+\frac{4\eta^{\prime}}{\omega_m}\right)^{-\frac14}.
\end{eqnarray}

In the interaction picture with respect to the free parts $\tilde{\omega}_m^{\prime}b^{\dag}b+\Delta_{\mathrm{eff}}c^{\dag}c$, $H_{\mathrm{eff}}^{\prime}$ in Eq.~(\ref{Eq21}) is transformed to
\begin{eqnarray}\label{Eq23}
\tilde{H}_{\mathrm{eff}}^{\prime}=G_{\mathrm{eff}}^{\prime}
\left[e^{-i(\tilde{\omega}_m^{\prime}-\Delta_{\mathrm{eff}})t}bc^{\dag}+e^{i(\tilde{\omega}_m^{\prime}-\Delta_{\mathrm{eff}})t}b^{\dag}c+
e^{-i(\tilde{\omega}_m^{\prime}+\Delta_{\mathrm{eff}})t}bc+e^{i(\tilde{\omega}_m^{\prime}+\Delta_{\mathrm{eff}})t}b^{\dag}c^{\dag}\right].
\end{eqnarray}
In Fig.~\ref{Fig6}, we show the energy-level diagram of above Hamiltonian in the resonant condition $\Delta_{\mathrm{eff}}=\tilde{\omega}_m^{\prime}$ clearly. We find that the cooling of mechanical mode corresponding to the anti-Stokes process can be significantly enhanced due to the resonant interaction. While the heating corresponding to the Stokes process is strongly suppressed since the detuning $2\tilde{\omega}_m^{\prime}$ is much larger than the coupling strength $G_{\mathrm{eff}}^{\prime}$ (in the present parameter regime, $2\tilde{\omega}_m^{\prime}/G_{\mathrm{eff}}^{\prime}\simeq32$). The resonant condition $\Delta_{\mathrm{eff}}=\tilde{\omega}_m^{\prime}=\omega_m\sqrt{1+\frac{4\eta^{\prime}}{\omega_m}}\simeq1.4\omega_m$ just is the optimal effective detuning $\Delta_{\mathrm{eff}}$ in Fig.~\ref{Fig5}.

\begin{figure}[H]
\centering
\includegraphics[scale=0.5]{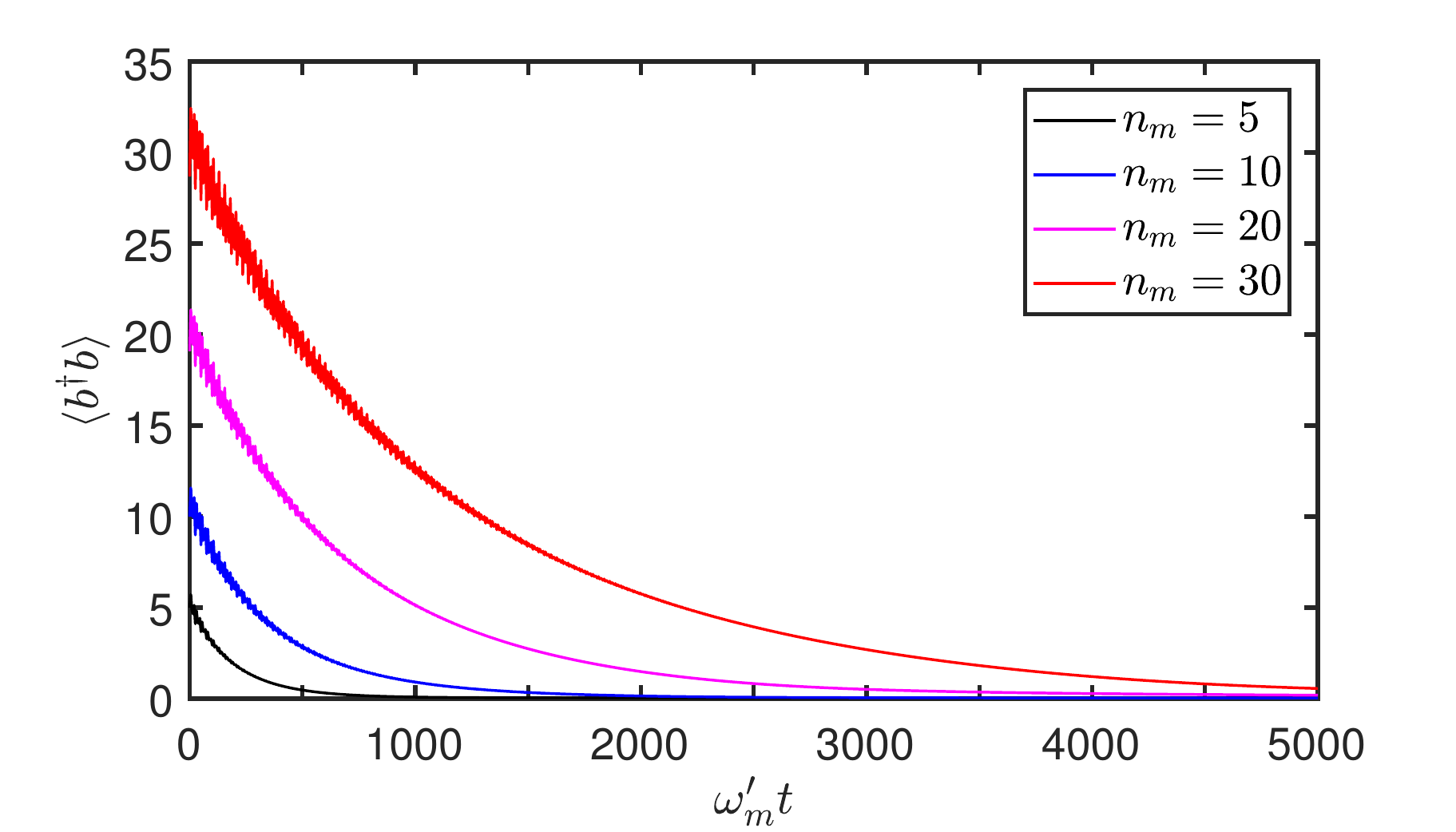}
\caption{(Color online) Time evolution of the mean phonon number $\langle b^{\dag}b\rangle$ corresponding to the optimal effective detuning $\Delta_{\mathrm{eff}}=1.4\omega_m$ when the initial state of the mechanical oscillator is a thermal state with certain mean thermal phonon number $n_m$. The other system parameters are the same as those in Fig.~\ref{Fig2}.}\label{Fig7}
\end{figure}

To further check the cooling effect of mechanical mode in the optimal effective detuning condition, there is necessary to present the evolution of the mean phonon number. Since the mechanical oscillator is initially at thermal equilibrium with its environment, it is physically reasonable to assume the mechanical oscillator is prepared in the initial thermal state $\varrho(0)=\sum_n(n_m)^n/(n_m+1)^{n+1}|n\rangle\langle n|$ with certain mean thermal phonon number $n_m$. Here $|n\rangle$ is the Fock basis. In Fig.~\ref{Fig7}, we plot the time evolution of the mean phonon number $\langle b^{\dag}b\rangle$ corresponding to the optimal effective detuning when the initial state of the mechanical oscillator is a thermal state. One can see that the final mean phonon number can be less than 1, therefore, which provides a prerequisite for the reveal of squeezing effect.

\subsection{The approach of covariance matrix}
\subsubsection{Dynamical equation for covariance matrix}
According to Eq.~(\ref{Eq2}), the set of nonlinear QLEs with system operators $q$, $p$, $a$, and $c$ is
\begin{eqnarray}\label{Eq24}
\frac{dq}{dt}&=&\omega_mp, \cr\cr
\frac{dp}{dt}&=&-(\omega_m+4\eta)q-\gamma_mp+g_0a^{\dag}a+\xi(t), \cr\cr
\frac{da}{dt}&=&-(\kappa+i\delta_c)a+ig_0aq-iGc-iE+\sqrt{2\kappa}a_{\mathrm{in}}(t), \cr\cr
\frac{dc}{dt}&=&-(\gamma_a+i\Delta_a)c-iGa+\sqrt{2\gamma_a}c_{\mathrm{in}}(t),
\end{eqnarray}
where $\xi(t)$ is the stochastic Hermitian Brownian noise operator which describes the dissipative friction forces subjecting to the mechanical oscillator. Its non-Markovian correlation function is given by~\cite{2001PRA63023812}
\begin{eqnarray}\label{Eq25}
\langle\xi(t)\xi(t^{\prime})\rangle=\frac{\gamma_m}{2\pi\omega_m}\int\omega\Big[\coth\Big(\frac{\hbar\omega}{2k_BT}\Big)+1\Big]e^{-i\omega(t-t^{\prime})}d\omega.
\end{eqnarray}
However, as to the case of $\omega_m\gg\gamma_m$ (a high quality mechanical oscillator), only the resonant noise components at frequency $\omega\sim\omega_m$ dominantly affect the dynamics of the mechanical oscillator. Thus the colored spectrum of Eq.~(\ref{Eq25}) can be simplified as the Markovian process and the correlation function becomes
\begin{eqnarray}\label{Eq26}
\langle\xi(t)\xi(t^{\prime})+\xi(t^{\prime})\xi(t)\rangle\simeq2\gamma_m\coth\Big(\frac{\hbar\omega_m}{2k_BT}\Big)\delta(t-t^{\prime})=
2\gamma_m(2n_m+1)\delta(t-t^{\prime}).
\end{eqnarray}

Exploiting above similar linearization procedure, the equation of motion corresponding to the classical mean values about $\langle q(t)\rangle$, $\langle p(t)\rangle$, $\langle a(t)\rangle$, and $c(t)$ is given by
\begin{eqnarray}\label{Eq27}
\langle\dot{q}(t)\rangle&=&\omega_m\langle p(t)\rangle, \cr\cr
\langle\dot{p}(t)\rangle&=&-(\omega_m+4\eta)\langle q(t)\rangle-\gamma_m\langle p(t)\rangle+g_0|\langle a(t)\rangle|^2, \cr\cr
\langle\dot{a}(t)\rangle&=&-(\kappa+i\delta_c)\langle a(t)\rangle+ig_0\langle a(t)\rangle\langle q(t)\rangle-iG\langle c(t)\rangle-iE, \cr\cr
\langle\dot{c}(t)\rangle&=&-(\gamma_a+i\Delta_a)\langle c(t)\rangle-iG\langle a(t)\rangle,
\end{eqnarray}
and the set of linearized QLEs for the quantum fluctuation operators $\delta q(t)$, $\delta p(t)$, $\delta a(t)$, and $\delta c(t)$ is
\begin{eqnarray}\label{Eq28}
\delta\dot{q}&=&\omega_m\delta p, \cr\cr
\delta\dot{p}&=&-(\omega_m+4\eta)\delta q-\gamma_m\delta p+g_0\langle a(t)\rangle^{\ast}\delta a+g_0\langle a(t)\rangle\delta a^{\dag}+\xi(t), \cr\cr
\delta\dot{a}&=&-[\kappa+i(\delta_c-g_0\langle q(t)\rangle)]\delta a+ig_0\langle a(t)\rangle\delta q-iG\delta c+\sqrt{2\kappa}a_{\mathrm{in}}(t), \cr\cr
\delta\dot{c}&=&-(\gamma_a+i\Delta_a)\delta c-iG\delta a+\sqrt{2\gamma_a}c_{\mathrm{in}}(t).
\end{eqnarray}

By introducing the quadrature operators for the cavity field, atoms, and their input noises:
\begin{eqnarray}\label{Eq29}
\delta x_1&=&(\delta a+\delta a^{\dag})/\sqrt{2},~~~~~~~~~~\delta y_1=(\delta a-\delta a^{\dag})/\sqrt{2}i, \cr\cr
\delta x_2&=&(\delta c+\delta c^{\dag})/\sqrt{2},~~~~~~~~~~\delta y_2=(\delta c-\delta c^{\dag})/\sqrt{2}i, \cr\cr
\delta x_1^{\mathrm{in}}&=&(a_{\mathrm{in}}+a_{\mathrm{in}}^{\dag})/\sqrt{2},~~~~~~~~~~
\delta y_1^{\mathrm{in}}=(a_{\mathrm{in}}-a_{\mathrm{in}}^{\dag})/\sqrt{2}i, \cr\cr
\delta x_2^{\mathrm{in}}&=&(c_{\mathrm{in}}+c_{\mathrm{in}}^{\dag})/\sqrt{2},~~~~~~~~~~
\delta y_2^{\mathrm{in}}=(c_{\mathrm{in}}-c_{\mathrm{in}}^{\dag})/\sqrt{2}i,
\end{eqnarray}
and the vectors of all quadratures and noises:
\begin{eqnarray}\label{Eq30}
U&=&[\delta q, \delta p, \delta x_1, \delta y_1, \delta x_2, \delta y_2]^T, \cr\cr
N&=&[0, \xi(t), \sqrt{2\kappa}\delta x_1^{\mathrm{in}}, \sqrt{2\kappa}\delta y_1^{\mathrm{in}}, \sqrt{2\gamma_a}\delta x_2^{\mathrm{in}}, \sqrt{2\gamma_a}\delta y_2^{\mathrm{in}}]^T,
\end{eqnarray}
the linearized QLEs for the quantum fluctuation operators in Eq.~(\ref{Eq28}) can be rewritten as
\begin{eqnarray}\label{Eq31}
\frac{dU}{dt}=A(t)U+N(t),
\end{eqnarray}
where $A(t)$ is a 6$\times$6 time-dependent matrix
\begin{eqnarray}\label{Eq32}
A(t)=
\begin{bmatrix}
0~~ & \omega_m~~ & 0~~ & 0~~ & 0~~ & 0~~ \\
-(\omega_m+4\eta) & -\gamma_m~~ & G_x(t)~~ & G_y(t)~~ & 0~~ & 0~~ \\
-G_y(t)~~ & 0~~ & -\kappa~~ & \Delta_c(t)~~ & 0~~ & G~~ \\
G_x(t)~~ & 0~~ & -\Delta_c(t)~~ & -\kappa~~ & -G~~ & 0~~ \\
0~~ & 0~~ & 0~~ & G~~ & -\gamma_a~~ & \Delta_a~~ \\
0~~ & 0~~ & -G~~ & 0~~ & -\Delta_a~~ & -\gamma_a
\end{bmatrix}.
\end{eqnarray}
Here, $\Delta_c(t)=\delta_c-g_0\langle q(t)\rangle$ is the effective time-modulated detuning and $G_x(t)$ and $G_y(t)$ are, respectively, the real and imaginary parts of the effective optomechanical coupling $G_0(t)=\sqrt{2}g_0\langle a(t)\rangle$. 

Due to the above linearized dynamics and the zero-mean Gaussian nature for the quantum noises, the quantum fluctuations in the stable regime will evolve to an asymptotic Gaussian state which can be characterized by the $6\times6$ covariance matrix completely
\begin{eqnarray}\label{Eq33}
V_{k,l}=\langle U_k(t)U_l(t)+U_l(t)U_k(t)\rangle/2.
\end{eqnarray}
From Eqs.~(\ref{Eq31}) and (\ref{Eq33}), we can deduce the dynamical equation which governs the evolution of the covariance matrix
\begin{eqnarray}\label{Eq34}
\dot{V}(t)=A(t)V(t)+V(t)A^T(t)+D,
\end{eqnarray}
where $A^T(t)$ denotes the transpose of $A(t)$ and $D=\mathrm{Diag}[0, \gamma_m(2n_m+1), \kappa, \kappa, \gamma_a, \gamma_a]$ is the matrix of noise correlation. Equation~(\ref{Eq34}) is an inhomogeneous first-order differential equation with 21 elements which can be numerically solved with the initial condition of covariance matrix $V(0)=\mathrm{Diag}[n_m+1/2, n_m+1/2, 1/2, 1/2, 1/2, 1/2]$.

\subsubsection{Time evolution of variance for the mirror position}
\begin{figure}
\centering
\includegraphics[scale=0.5]{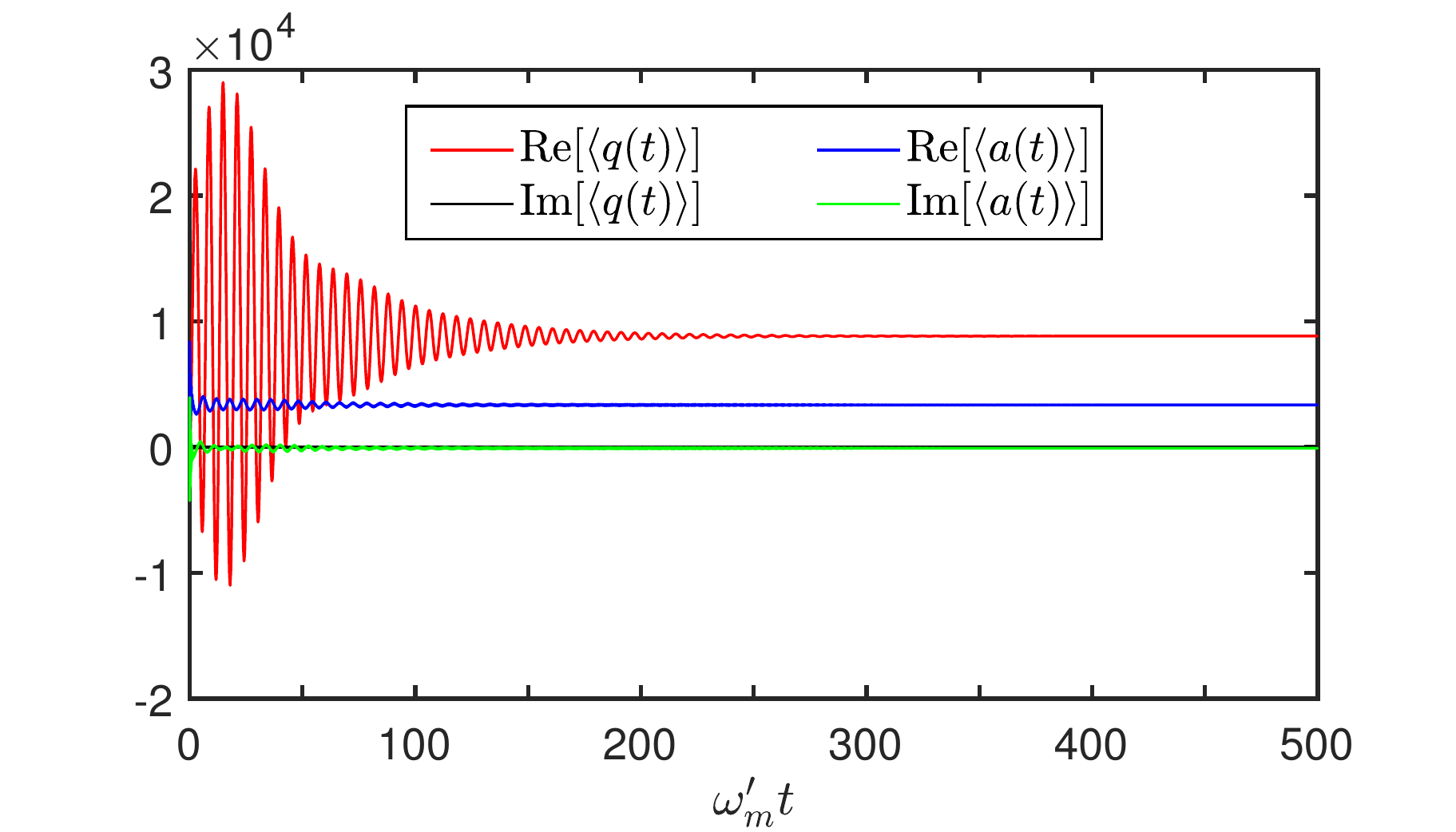}
\caption{(Color online) Time evolution of the real and imaginary parts of the mirror position mean value $\langle q(t)\rangle$ and the cavity mode mean value $\langle a(t)\rangle$. The parameters are the same as those in Fig.~\ref{Fig2}.}\label{Fig8}
\end{figure}
In Fig.~\ref{Fig8}, we show the time evolution of the real and imaginary parts of the mirror position mean value $\langle q(t)\rangle$ and the cavity mode mean value $\langle a(t)\rangle$. We find that the real and imaginary parts of the mean values reach the steady states quickly and the real part is much larger than the imaginary part ($\mathrm{Re}[\langle q(t)\rangle]\gg\mathrm{Im}[\langle q(t)\rangle]$ and $\mathrm{Re}[\langle a(t)\rangle]\gg\mathrm{Im}[\langle a(t)\rangle]$). Thus, we can make the approximations as above subsection
\begin{eqnarray}\label{Eq35}
\langle q(t)\rangle\simeq|\langle q\rangle_s|,~~~~~~~~~~\langle a(t)\rangle\simeq\langle a(t)\rangle^{\ast}\simeq|\langle a\rangle_s|,
\end{eqnarray}
where $|\langle q\rangle_s|$ and $|\langle a\rangle_s|$ denote the steady state mean values of the mirror position and cavity mode, respectively.

\begin{figure}
\centering
\includegraphics[scale=0.5]{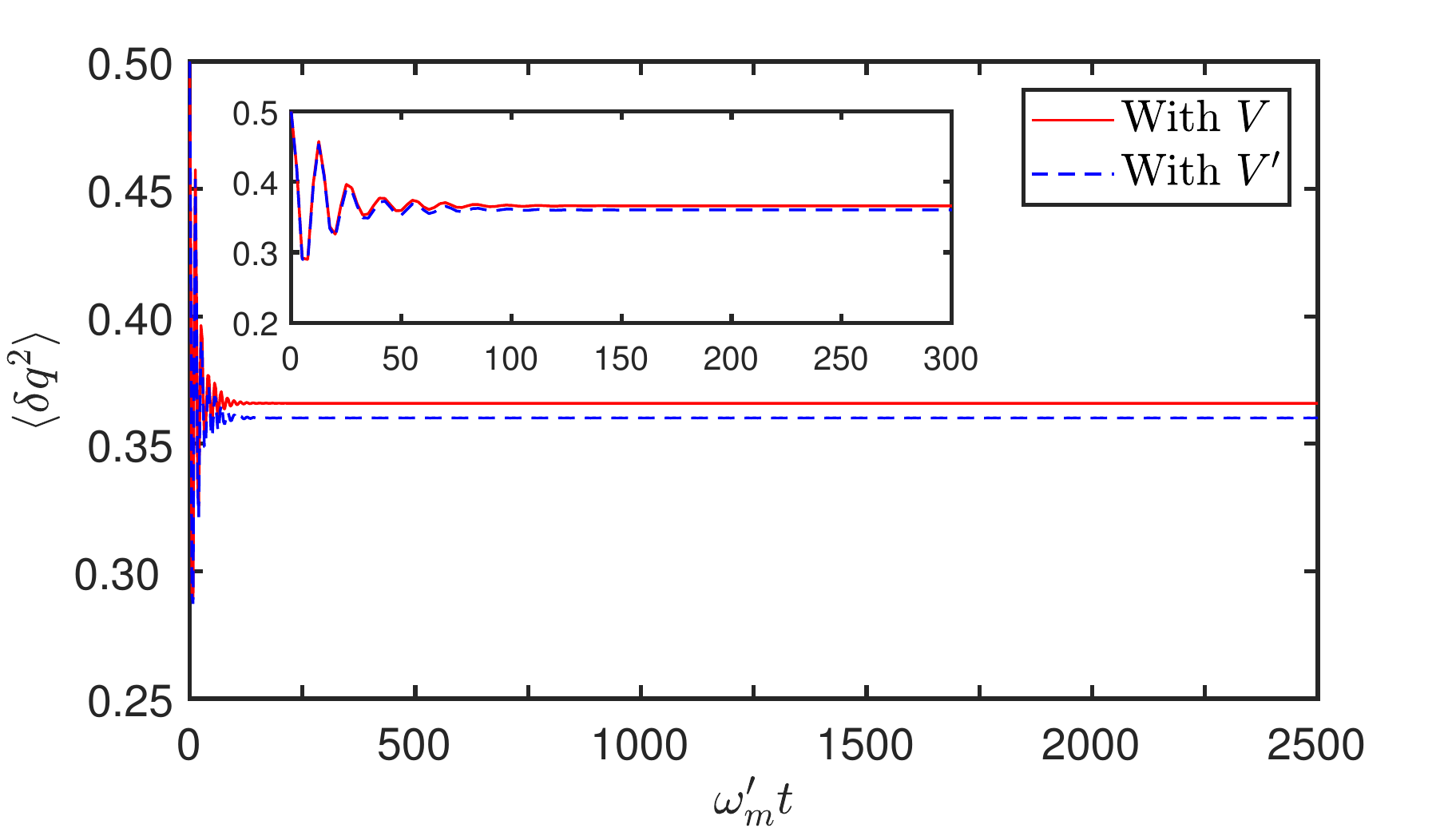}
\caption{(Color online) Time evolution of the variance $\langle\delta q^2\rangle$ for the mirror position. The red solid line and blue dashed line denote, respectively, the corresponding numerical result with full covariance matrix $V$ and reduced covariance matrix $V^{\prime}$. The system parameters are the same as those in Fig.~\ref{Fig2}. Here the mean thermal phonon number has been set $n_m=0$.}\label{Fig9}
\end{figure}

By numerically solving the dynamical equation about covariance matrix $V$ in Eq.~(\ref{Eq34}) under above approximations, we plot the time evolution of variance $\langle\delta q^2\rangle$ for the mirror position in Fig.~\ref{Fig9}. From Fig.~\ref{Fig9}, one notes that the variance $\langle\delta q^2\rangle$ also finally reaches its steady-state value below $1/2$.

In fact, exploiting the similar means of eliminating the cavity mode adiabatically, we can obtain the dynamical equation about the reduced covariance matrix $V^{\prime}$:
\begin{eqnarray}\label{Eq36}
\dot{V}^{\prime}(t)=BV^{\prime}(t)+V^{\prime}(t)B^T+D^{\prime},
\end{eqnarray}
where
\begin{eqnarray}\label{Eq37}
B=
\begin{bmatrix}
~~~~~0~~~~~ & ~~~~~\omega_m~~~~~ & ~~~~~0~~~~~ & ~~~~~0~~~~~ \\
-\left(\omega_m+4\eta-\frac{2g_0^2|\langle a\rangle_s|^2}{\Delta_c}\right) & -\gamma_m & -\frac{\sqrt{2}g_0|\langle a\rangle_s|G}{\Delta_c} & 0 \\
0 & 0 & -\gamma_a & \Delta_a-\frac{G^2}{\Delta_c} \\
-\frac{\sqrt{2}g_0|\langle a\rangle_s|G}{\Delta_c} & 0 & -(\Delta_a-\frac{G^2}{\Delta_c}) & -\gamma_a
\end{bmatrix},
\end{eqnarray}
and $D^{\prime}=\mathrm{Diag}[0, \gamma_m(2n_m+1), \gamma_a, \gamma_a]$.

To check the validity of the dynamical equation about the reduced covariance matrix $V^{\prime}$ in Eq.~(\ref{Eq36}), we also present the time evolution of variance $\langle\delta q^2\rangle$ in Fig.~\ref{Fig9}. Compared with the result obtained from the full covariance matrix $V$, the result obtained from the reduced covariance matrix $V^{\prime}$ is agreed very well.

\subsubsection{Variance for the mirror position in the steady-state regime}
When the system reaches the steady state, the reduced covariance matrix $V^{\prime}$ is dominated by the following Lyapunov equation
\begin{eqnarray}\label{Eq38}
BV^{\prime}+V^{\prime}B^T=-D^{\prime}.
\end{eqnarray}
Equation~(\ref{Eq38}) can be analytically solved in the parameter regime with the negligible mechanical damping $\gamma_m\simeq0$ and the variance $\langle\delta q^2\rangle$ for the mirror position in the steady state is given by
\begin{eqnarray}\label{Eq39}
\langle\delta q^2\rangle\simeq
\left[\omega_m+\frac{(\Delta_G^2+\gamma_a^2)^2}{\Omega_m(\Delta_G^2+\gamma_a^2)-G_g^2\Delta_G}\right]/4\Delta_G,
\end{eqnarray}
in which
\begin{eqnarray}
\Delta_G&=&\Delta_a-\frac{G^2}{\Delta_c}, ~~~~~~~~
\Omega_m=\omega_m+4\eta-\frac{2g_0^2|\langle a\rangle_s|^2}{\Delta_c}, ~~~~~~~~
G_g=\frac{\sqrt{2}g_0|\langle a\rangle_s|G}{\Delta_c}.
\end{eqnarray}

\begin{figure}
\centering
\includegraphics[scale=0.5]{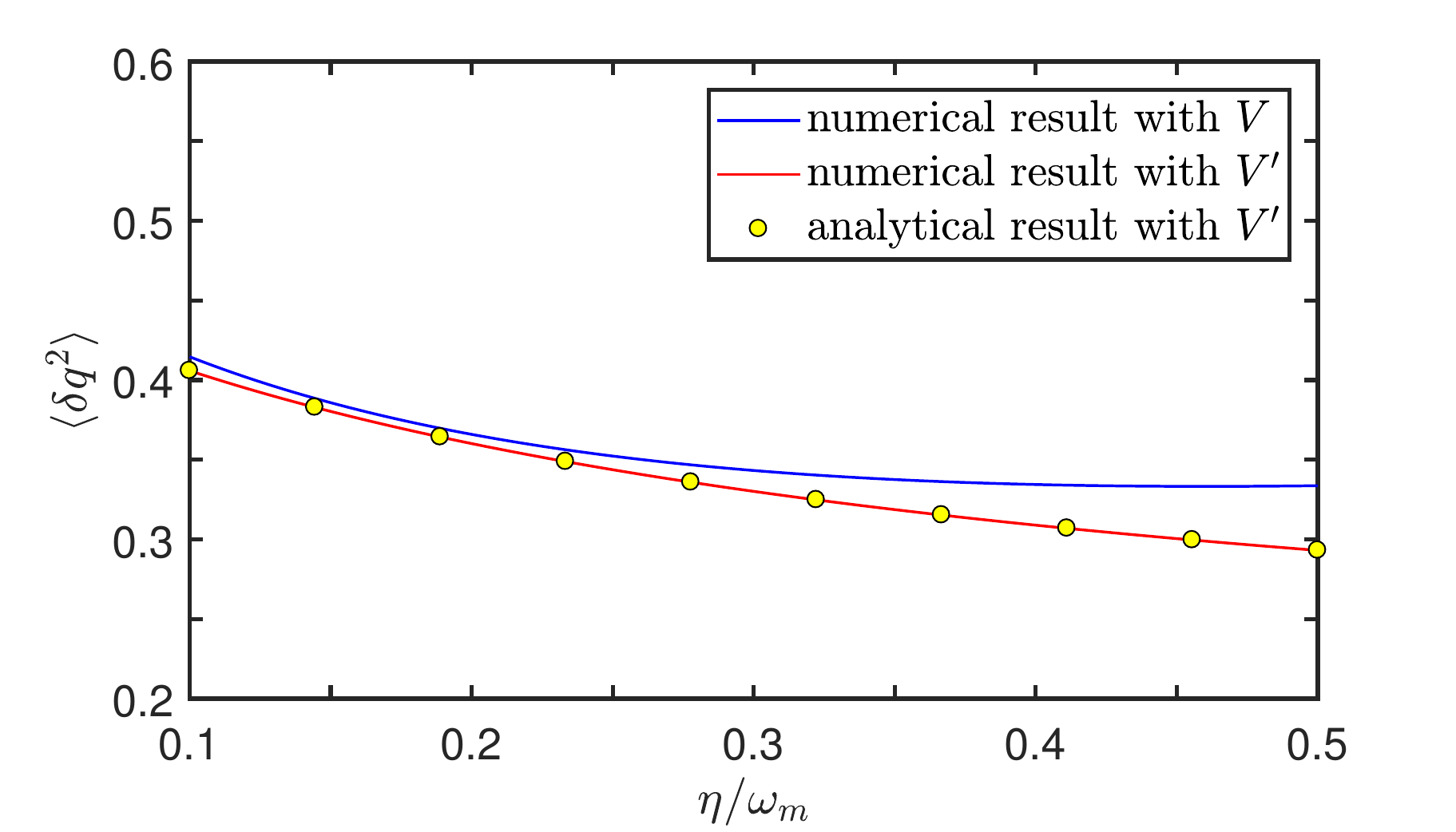}
\caption{(Color online) Steady-state variance $\langle\delta q^2\rangle$ for the mirror position versus $\eta$. The blue line, red line, and yellow dots indicate, respectively, the numerical result with full covariance matrix $V$, numerical result with reduced covariance matrix $V^{\prime}$, and analytical result with reduced covariance matrix $V^{\prime}$. The other parameters are the same as those in Fig.~\ref{Fig2}.}\label{Fig10}
\end{figure}

The steady-state variance $\langle\delta q^2\rangle$ for the mirror position obtained from, respectively, numerical result with full covariance matrix $V$, numerical result with reduced covariance matrix $V^{\prime}$, and analytical result with reduced covariance matrix $V^{\prime}$, versus $\eta$ is shown in Fig.~\ref{Fig10}. We can find that, as to the result obtained from the reduced covariance matrix $V^{\prime}$, the numerical and analytical solutions are agreed very well. In the appropriate parameter scale of $\eta$, the results obtained from the full covariance matrix $V$ and reduced covariance matrix $V^{\prime}$ are also agreed well. However, with the gradual increase of $\eta$, the two results begin to exist the difference and no longer agree well each other. This is because eliminating the cavity mode adiabatically is valid only in the suitable parameter regime of $\eta$.

\subsection{Comparison of the two approaches}
In the above subsections, we discuss the squeezing of the movable mirror based on the approaches of master equation and covariance matrix in detail, respectively. In this subsection, we make a comparison for the two different approaches.

In the approach of master equation, we make the approximations: $\langle a(t)\rangle\simeq\langle a(t)\rangle^{\ast}\simeq|\langle a\rangle_s|$ and $\langle b(t)\rangle\simeq\langle b(t)\rangle^{\ast}\simeq|\langle b\rangle_s|$. In fact, the Hamiltonian of the time-dependent dynamics for the system is
\begin{eqnarray}\label{Eq41}
H_{\mathrm{lin}}^{\prime}&=&
[\delta_c-g_0^{\prime}(\langle b(t)\rangle+\langle b(t)\rangle^{\ast})]a^{\dag}a+\omega_m^{\prime}b^{\dag}b+\Delta_ac^{\dag}c+\eta(b^2+b^{\dag2})+ \cr\cr
&&G(c^{\dag}a+ca^{\dag})-g_0^{\prime}\langle a(t)\rangle^{\ast}a(b+b^{\dag})-g_0^{\prime}\langle a(t)\rangle a^{\dag}(b+b^{\dag}),
\end{eqnarray}
where $\langle a(t)\rangle$ and $\langle b(t)\rangle$ are the solutions of the set of nonlinear differential equations in Eq.~(\ref{Eq7}). Similarly, in the approach of covariance matrix, we also make the approximations； $\langle q(t)\rangle\simeq|\langle q\rangle_s|$ and $\langle a(t)\rangle\simeq\langle a(t)\rangle^{\ast}\simeq|\langle a\rangle_s|$.
\begin{figure}
\centering
\includegraphics[scale=0.5]{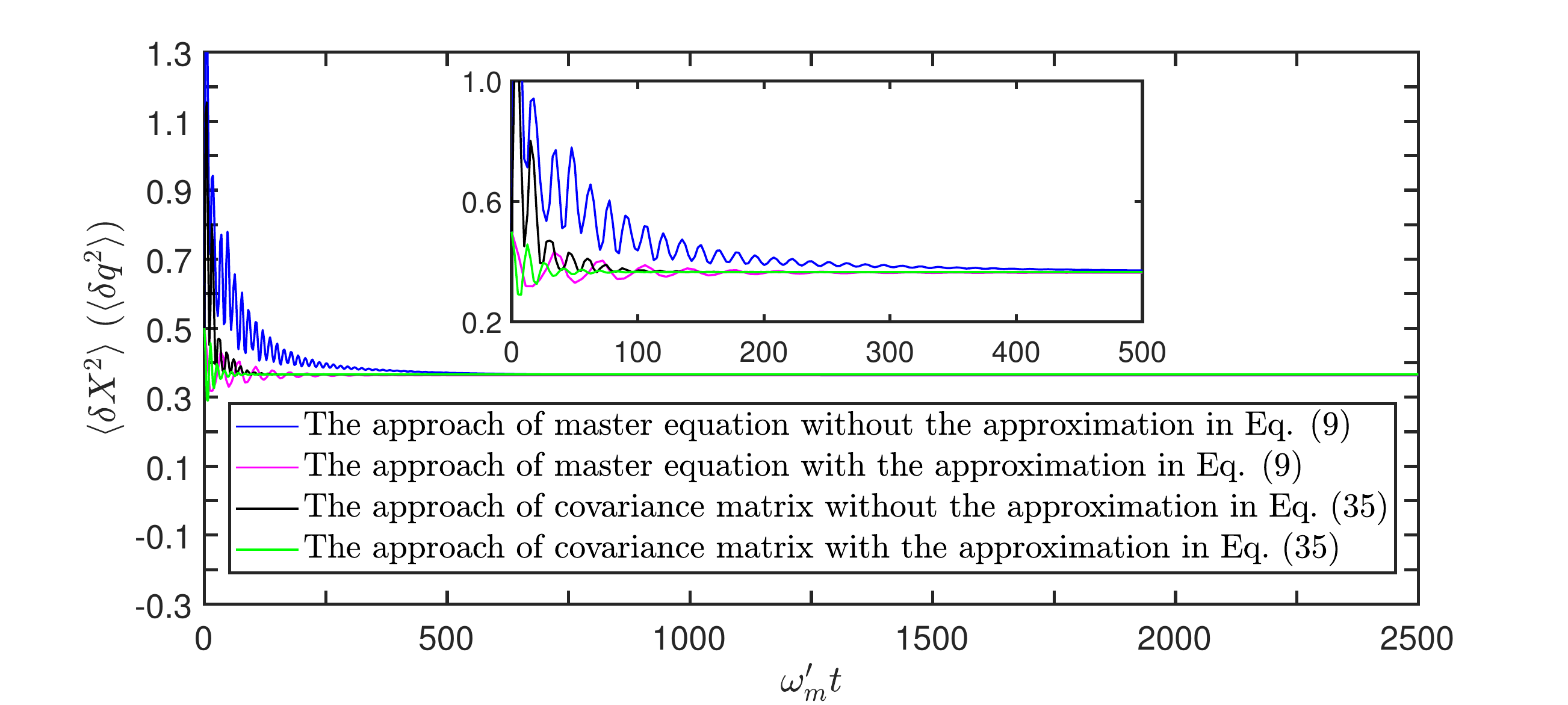}
\caption{(Color online) Time evolution of variance $\langle\delta X^2\rangle$ for the quadrature operator $X$ (variance $\langle\delta q^2\rangle$ for the mirror position) obtained from the master equation (dynamical equation for covariance matrix) in the both cases of without and with approximations. The system parameters are the same as those in Fig.~\ref{Fig2}.}\label{Fig11}
\end{figure}

To further show the exact dynamics of system explicitly and compare the results obtained from above two different approaches clearly, we numerically solve the master equation in Eq.~(\ref{Eq16}) with the time-dependent Hamiltonian in Eq.~(\ref{Eq41}) and dynamical equation for covariance matrix in Eq.~(\ref{Eq34}) with $6\times6$ time-dependent matrix $A(t)$ in Eq.~(\ref{Eq32}) once again. In Fig.~\ref{Fig11}, we present the time evolution of variance $\langle\delta X^2\rangle$ for the quadrature operator $X$ (variance $\langle\delta q^2\rangle$ for the mirror position) obtained from the master equation (dynamical equation for covariance matrix) in the both cases of without and with approximations. One can note that, as to the two different approaches, although they have different oscillation modes before reaching the steady state, they will reach the same steady state in the long enough time. And as to the cases of with approximations, they all obtain the steady state which is same with the exact dynamics in the long-time limit. Therefore, the approaches of master equation and dynamical equation for covariance matrix are completely equivalent in stationary behavior but differ in dynamical behavior. In terms of stationary
behavior, simplifying the system dynamics with the approximations of Eqs.~(\ref{Eq9}) and (\ref{Eq35}) is very reasonable.

\section{Implementation with a circuit-QED setup}\label{Sec4}
In this section, we give some brief discussions about the implementation of the present scheme with circuit-QED systems. As we all know, the circuit-QED system is formed by the superconducting flux qubit and the integrated-element LC oscillator~\cite{2017NP134447}. Additionally, the generation of nonclassical states and the sensitive detection of the position about the quantized mechanical motion have also been studied based on the superconducting circuit system both in theory and experiment~\cite{2016SB61163,2008NaturePhysics4785}.
In the present scheme, the generation of the second-order term introduced in Hamiltonian (\ref{Eq1}) is significantly important for realizing the mechanical squeezing, which is associated with the coupling between the oscillator and qubit. In fact, this oscillator-qubit coupling can be effectively implemented in circuit-QED system~\cite{2009Nature459960,2018PRA98023821}. As to the method of generation of the additional section-order term, the section of appendix in Ref.~\cite{2018AOP39239} has discussed in detail. In addition, as to other couplings involved in our scheme (coupling between atom and cavity and coupling between cavity mode and mechanical mode), we can exploit the similar architecture of superconducting circuits in Ref.~\cite{2016PRL116233604} to realize.

\section{Conclusions}\label{Sec5}
In conclusion, based on the master equation and covariance matrix these two different approaches, we have detailedly investigated the squeezing effect of the mirror motion in a hybrid system consisting of an atomic ensemble trapped inside a dissipative optomechanical cavity assisted with the perturbative oscillator coupling. We find that adiabatic elimination of the highly dissipative cavity mode to significantly simplify the system dynamics in both approaches is very effective. In the approach of master equation, from the effective Hamiltonian
Eqs.~(\ref{Eq14}) and (\ref{Eq23}), we numerically and dynamically derive the optimal effective detuning, respectively. Under the optimal effective detuning condition, we also check the cooling effects of the mechanical mode when the mechanical oscillator is initially prepared in a thermal state with certain mean thermal phonon number and find that the mechanical mode can be cooled down close to its ground state in the long-time limit, which provides a prerequisite for the generation of stationary squeezing. As to the covariance matrix approach, we reduce the dynamical equation of $6\times6$ covariance matrix as the one of $4\times4$ covariance matrix by eliminating the highly dissipative cavity mode adiabatically, which greatly simplifies the system dynamics. In this case, we obtain the analytical solution of the steady-state variance for the mechanical position approximately. Finally, we make a clear comparison for these two different approaches and find that they are completely equivalent for the stationary dynamics. The present scheme may be implemented with the circuit-QED systems and benefit forward the possible ultraprecise quantum measurement involved mechanical squeezing.

\begin{center}
$\mathbf{Acknowledgements}$
\end{center}

This work was supported by the National Natural Science Foundation of China under Grant Nos.
61822114, 11465020, 61465013; The Project of Jilin Science and Technology Development for Leading Talent of Science and Technology Innovation in Middle and
Young and Team Project under Grant (20160519022JH).

\appendix
\section{Adiabatic elimination of cavity mode}\label{App1}
Under the approximation of Eq.~(\ref{Eq9}), the linearized QLEs Eq.~(\ref{Eq8}) can be simplified as
\begin{eqnarray}\label{Ap1}
\dot{a}&=&-(\kappa+i\Delta_c)a-iGc+iG_0(b+b^{\dag})+\sqrt{2\kappa}a_{\mathrm{in}}(t), \cr\cr
\dot{b}&=&-(\gamma_m+i\omega_m^{\prime})b+iG_0(a+a^{\dag})-2i\eta b^{\dag}+\sqrt{2\gamma_m}b_{\mathrm{in}}(t),\cr\cr
\dot{c}&=&-(\gamma_a+i\Delta_a)c-iGa+\sqrt{2\gamma_a}c_{\mathrm{in}}(t),
\end{eqnarray}
where $\Delta_c=\delta_c-2g_0^{\prime}|\langle b\rangle_s|$ and $G_0=g_0^{\prime}|\langle a\rangle_s|$. The formal solution of Eq.~(\ref{Ap1}) is
\begin{eqnarray}\label{Ap2}
a(t)&=&a(0)e^{-(\kappa+i\Delta_c)t}+e^{-(\kappa+i\Delta_c)t}\int_0^t\left\{-iGc(\tau)+iG_0[b(\tau)+b^{\dag}(\tau)]+\sqrt{2\kappa}a_{\mathrm{in}}(\tau)\right\}e^{(\kappa+i\Delta_c)\tau}d\tau, \cr\cr
b(t)&=&b(0)e^{-(\gamma_m+i\omega_m^{\prime})}t+e^{-(\gamma_m+i\omega_m^{\prime})t}\int_0^t\left\{iG_0[a(\tau)+a^{\dag}(\tau)]-2i\eta b^{\dag}(\tau)+\sqrt{2\gamma_m}b_{\mathrm{in}}(\tau)\right\}e^{(\gamma_m+i\omega_m^{\prime})\tau}d\tau, \cr\cr
c(t)&=&c(0)e^{-(\gamma_a+i\Delta_a)t}+e^{-(\gamma_a+i\Delta_a)t}\int_0^t\left[-iGa(\tau)+\sqrt{2\gamma_a}c_{\mathrm{in}}(\tau)\right]e^{(\gamma_a+i\Delta_a)\tau}d\tau.
\end{eqnarray}

When the decay rate of cavity field is much larger than the damping rate of mechanical oscillator and the decay rate of atoms, the cavity mode $a$ can only slightly affect the dynamics of mechanical mode $b$ and atom mode $c$. So the approximate expressions about modes $b$ and $c$ are
\begin{eqnarray}\label{Ap3}
b(t)&\simeq&b(0)e^{-(\gamma_m+i\omega_m^{\prime})t}+B_{\mathrm{in}}^{\prime}(t), \cr\cr
c(t)&\simeq&c(0)e^{-(\gamma_a+i\Delta_a)t}+C_{\mathrm{in}}^{\prime}(t),
\end{eqnarray}
where $B_{\mathrm{in}}^{\prime}(t)$ and $C_{\mathrm{in}}^{\prime}(t)$ represent the noise terms. By substituting Eq.~(\ref{Ap3}) into the expression about mode $a$ in Eq.~(\ref{Ap2}), we obtain
\begin{eqnarray}\label{Ap4}
a(t)&\simeq&a(0)e^{-(\kappa+i\Delta_c)t}+
e^{-(\kappa+i\Delta_c)t}\int_0^t\Big\{-iGc(0)e^{-(\gamma_a+i\Delta_a)\tau}+ \cr\cr
&&iG_0\big[b(0)e^{-(\gamma_m+i\omega_m^{\prime})\tau}+b^{\dag}(0)e^{-(\gamma_m-i\omega_m^{\prime})\tau}\big]\Big\}e^{(\kappa+i\Delta_c)\tau}d\tau+A_{\mathrm{in}}(t) \cr\cr
&=&a(0)e^{-(\kappa+i\Delta_c)t}+\frac{-iGc(0)e^{-(\gamma_a+i\Delta_a)t}}{(\kappa-\gamma_a)+i(\Delta_c-\Delta_a)}-
\frac{-iGc(0)e^{-(\kappa+i\Delta_c)t}}{(\kappa-\gamma_a)+i(\Delta_c-\Delta_a)}+ \cr\cr
&&\frac{iG_0b(0)e^{-(\gamma_m+i\omega_m^{\prime})t}}{(\kappa-\gamma_m)+i(\Delta_c-\omega_m^{\prime})}-
\frac{iG_0b(0)e^{-(\kappa+i\Delta_c)t}}{(\kappa-\gamma_m)+i(\Delta_c-\omega_m^{\prime})}+ \cr\cr
&&\frac{iG_0b^{\dag}(0)e^{-(\gamma_m-i\omega_m^{\prime})t}}{(\kappa-\gamma_m)+i(\Delta_c+\omega_m^{\prime})}-
\frac{iG_0b^{\dag}(0)e^{-(\kappa+i\Delta_c)t}}{(\kappa-\gamma_m)+i(\Delta_c+\omega_m^{\prime})}+A_{\mathrm{in}}^{\prime}(t),
\end{eqnarray}
where $A_{\mathrm{in}}^{\prime}(t)$ denotes the noise term. In the parameter regimes $|\Delta_c|\gg(\omega_m^{\prime}, \Delta_a)$ and $\kappa\gg(\gamma_m, \gamma_a)$, Eq.~(\ref{Ap4}) can be simplified as
\begin{eqnarray}\label{Ap5}
a(t)&\simeq&a(0)e^{-(\kappa+i\Delta_c)t}+\frac{-iGc(0)e^{-(\gamma_a+i\Delta_a)t}}{(\kappa-\gamma_a)+i(\Delta_c-\Delta_a)}+ \cr\cr
&&\frac{iG_0b(0)e^{-(\gamma_m+i\omega_m^{\prime})t}}{(\kappa-\gamma_m)+i(\Delta_c-\omega_m^{\prime})}+
\frac{iG_0b^{\dag}(0)e^{-(\gamma_m-i\omega_m^{\prime})t}}{(\kappa-\gamma_m)+i(\Delta_c+\omega_m^{\prime})}+A_{\mathrm{in}}^{\prime}(t),
\end{eqnarray}
by using Eq.~(\ref{Ap3}) once again,
\begin{eqnarray}\label{Ap6}
a(t)&\simeq&a(0)e^{-(\kappa+i\Delta_c)t}+\frac{-iGc(t)}{(\kappa-\gamma_a)+i(\Delta_c-\Delta_a)}+ \cr\cr
&&\frac{iG_0b(t)}{(\kappa-\gamma_m)+i(\Delta_c-\omega_m^{\prime})}+
\frac{iG_0b^{\dag}(t)}{(\kappa-\gamma_m)+i(\Delta_c+\omega_m^{\prime})}+A_{\mathrm{in}}(t) \cr\cr
&\simeq&a(0)e^{-(\kappa+i\Delta_c)t}+\frac{-iGc(t)}{\kappa+i\Delta_c}+
\frac{iG_0[b(t)+b^{\dag}(t)]}{\kappa+i\Delta_c}+A_{\mathrm{in}}(t),
\end{eqnarray}
where $A_{\mathrm{in}}(t)$ is the modified noise operator. Since $\kappa$ is large, the term containing $\mathrm{exp}(-\kappa t)$ in Eq.~(\ref{Ap6}) is a fast decaying term and thus it can be neglected safely. Therefore cavity mode $a(t)$ now can be expressed in terms of $b(t)$, $b^{\dag}(t)$, and $c(t)$
\begin{eqnarray}\label{Ap7}
a(t)\simeq\frac{iG_0[b(t)+b^{\dag}(t)]}{\kappa+i\Delta_c}+\frac{-iGc(t)}{\kappa+i\Delta_c}+A_{\mathrm{in}}(t).
\end{eqnarray}

\section{Derivation of squeezing parameter}\label{App2}
Applying the squeezing transformation $S(r)=\mathrm{exp}\left[\frac r2(b^2-b^{\dag 2})\right]$ to the effective Hamiltonian $H_{\mathrm{eff}}$ in Eq.~(\ref{Eq14}), we obtain
\begin{eqnarray}\label{Ap8}
H_{\mathrm{eff}}^{\prime}&=&\left[\tilde{\omega}_m(\cosh^2r+\sinh^2r)-4\eta^{\prime}\cosh r\sinh r\right]b^{\dag}b+
\Delta_{\mathrm{eff}}c^{\dag}c+G_{\mathrm{eff}}(\cosh r-\sinh r)\times \cr\cr
&&(b+b^{\dag})(c+c^{\dag})+\left[-\tilde{\omega}_m\cosh r\sinh r+\eta^{\prime}(\cosh^2r+\sinh^2r)\right](b^2+b^{\dag2}).
\end{eqnarray}
By setting $-\tilde{\omega}_m\cosh r\sinh r+\eta^{\prime}(\cosh^2r+\sinh^2r)=0$, the squeezing parameter $r$ can be obtained
\begin{eqnarray}\label{Ap9}
r=\frac14\ln\left(1+\frac{4\eta^{\prime}}{\omega_m}\right).
\end{eqnarray}
Thus,
\begin{eqnarray}
\omega_m^{\prime}&=&\tilde{\omega}_m(\cosh^2r+\sinh^2r)-4\eta^{\prime}\cosh r\sinh r=\omega_m\sqrt{1+\frac{4\eta^{\prime}}{\omega_m}}, \cr\cr
G_{\mathrm{eff}}^{\prime}&=&G_{\mathrm{eff}}(\cosh r-\sinh r)=G_{\mathrm{eff}}\left(1+\frac{4\eta^{\prime}}{\omega_m}\right)^{-\frac14}.
\end{eqnarray}


\end{document}